\documentclass[preprint,eqsecnum,preprintnumbers,nofootinbib,byrevtex,prd,aps,showpacs,showkeys,groupedaddress,floatfix]{revtex4}
\usepackage{amsmath,amssymb} 
\usepackage[dvips]{graphicx} 
\usepackage{graphics}
\usepackage{graphicx}
\usepackage{epsfig}
\usepackage{slashed}
\usepackage[english]{babel}

\begin{document}


\title{Approximate Solutions to the Klein-Fock-Gordon Equation for the sum of Coulomb and Ring-Shaped like potentials}

\author{Sh. M. Nagiyev$^{1}$, A. I. Ahmadov$^{2,3}$ \footnote{Corresponding author: ahmadovazar@yahoo.com}, and V. A. Tarverdiyeva$^{1}$}

\noindent

\affiliation{$^{1}$ Institute of Physics, Azerbaijan National Academy of Sciences, H. Javid Avenue, 131, AZ-1143, Baku, Azerbaijan}

\affiliation{$^{2}$ Department of Theoretical Physics, Baku State University, Z. Khalilov St. 23, AZ-1148, Baku, Azerbaijan}

\affiliation{$^{3}$ Institute of Physical Problems, Baku State University, Z. Khalilov St. 23, AZ-1148, Baku, Azerbaijan}


\begin{abstract}
We consider the quantum mechanical problem of the motion of a spinless 
charged relativistic particle with mass$M$, described
 by the Klein-Fock-Gordon equation with equal scalar $S(\vec{r})$ and 
 vector $V(\vec{r})$ Coulomb plus ring-shaped potentials.
It is shown that the system under consideration has both a discrete
at $\left|E\right|<Mc^{2} $ and a continuous at
$\left|E\right|>Mc^{2} $ energy spectra. We find the analytical
expressions for the corresponding complete wave functions. A
dynamical symmetry group $SU(1,1)$ for the radial wave equation of
motion is constructed. The algebra of generators of this group makes
it possible to find energy spectra in a purely algebraic way. It is
also shown that relativistic expressions for wave functions, energy
spectra and group generators in the limit $c\to \infty $ go over
into the corresponding expressions for the nonrelativistic problem.


\end{abstract}

\maketitle


\section{Introduction}

Nonrelativistic Schr\"{o}dinger and relativistic Dirac,
Klein-Fock-Gordon (KFG), and finite-difference equations describe
the systems in the nuclear physics, elementary physics as well as
the in atomic and molecular physics.[1-6] Examples to the commonly
used potentials in these equations are the Coulomb potential  and
the harmonic oscillator potential, as well as their various
varieties. Other types of interaction potentials include such as the
Kratzer,[7] Morse,[8] Eckart,[9] Manning-Rosen,[10]
P\"{o}schl-Teller,[11] Hulth\`{e}n,[12] Wood-Saxon,[13]
Makarov,[14] Hartmann,[15] and Hautot potentials[16]. Exact
solvability and the application range of these potentials are
examples to the main properties of them. Several approaches were
developed to study the quantum systems either relativistic or
nonrelativistic with non-zero angular momentum.[17-23]

The numerous interesting works devoted to the study of KFG equation
with different potentials. The $s$-wave KFG equation with the vector
Hulth\`{e}n type potential is one of the examples which was
investigated by use of the standard methods. Afterwards, the further
study for the vector and scalar Hulth\`{e}n type potentials were
carried out by Adame et al.[24,25] The same potential for $s$-wave
KFG equation were obtained for regular and irregular boundary
conditions in Ref. [26]. The path integral approach to the Green
function for the KFG equation with these two potentials was studied
in Ref. [27].

In Ref. [28-33], the scalar potential is equal and not equal to the
vector potential were assumed to get the bound states of the KFG
equation with some typical potential by using the ordinary quantum
mechanics.
Furthermore, KFG equation with the ring-shaped like potential was
investigated by Dong et al. [32] If we consider the case where the
interaction potential is not enough to create particle-antiparticle pairs,
the KFG equation can be applied to the treatment of a zero-spin particle and
apply the Dirac equation to that of a 1/2-spin particle. When a particle is
 in a strong field, the relativistic wave equations should
be considered in the quantum system. In any case, we can make
the correction easily for non-relativistic quantum mechanics.

Since the Coulomb potential is one of the exactly solvable potentials
in physics, it has wide range of application to the several research areas
such as nuclear and particle, atomic, condensed matter,
and chemical physics. It is natural and interesting question to study
the relativistic effects for a particle within this potential especially for strong coupling. Notice that KFG equation cannot be
solved exactly for many potentials of interactions for nonzero angular momentum $l\ne 0$ -states because of the centrifugal term of potentials.

Noncentral potentials which play an important role for describing the quantum systems, have the following form in spherical coordinates
\begin{equation} \label{GrindEQ__1_1_}
V(r,\theta ,\varphi )=V(r)+\frac{f(\theta ,\varphi )}{r^{2} } ,
\end{equation}
where $V(r)$ and $f(\theta ,\varphi )$ are certain functions of their arguments.
Coulomb or harmonic oscillator potentials are frequently used for the central part of the equation \eqref{GrindEQ__1_1_}.
One of the convenient combinations is the Hartmann potential[15] for which
\begin{equation} \label{GrindEQ__1_2_}
V=\eta \sigma ^{2} \left(\frac{2a_{0} }{r} -\eta \frac{a_{0}^{2} }{r^{2} \sin ^{2} \theta } \right)\varepsilon _{0} ,
\end{equation}
here  $a_{0} $  is the Bohr radius, $a_{0} =\hbar ^{2} /me^{2} $, $\varepsilon _{0} $
is the ground state energy  of the hydrogen atom, $\varepsilon _{0} ==me^{4} /2\hbar ^{2} $, $\eta $ and $\sigma $
are positive real numbers.
This potential was proposed by Hartmann in the framework
 of the nonrelativistic quantum mechanics for describing the organic molecules such as
benzene within quantum chemistry   to describing organic molecules in quantum chemistry like benzene.
One can use the same potential in nuclear physics for studying the interactions between deformed nuclei.
Other significant noncentral potentials of the form
\eqref{GrindEQ__1_1_} were considered by Hautot [16] in the
framework of the nonrelativistic quantum mechanics to study the
problem of the motion of a charged particle.

The original Hautot potential is defined in the following form:

\begin{equation} \label{GrindEQ__1_3_}
V(r,\theta )=\frac{\mu }{e} \left[V(r)+\frac{f(\theta )}{r^{2} } \right],
\end{equation}
here $\mu $ is the mass of the charged particle, $e$ is the charge of the particle.

He considered the two- and three-dimensional harmonic oscillator potentials and the
Coulomb potential to which terms of the type were added and found such functions that
enabled him to solve exactly the corresponding Schrödinger equations. Hence the Hartmann
potential is a special case of the Hautot potentials.~

One needs to study the noncentral potentials rather than the central ones for getting
better results in molecular structures and interactions. Examples to these situations
include the use of the ring-shaped potentials in quantum chemistry to describe the ring-shaped
organic molecules and in nuclear physics to investigate the interaction between the deformed
nucleus and spin-orbit coupling for a motion of the particle in the potential fields.
This noncentral potential is also used as a mathematical model in the description of diatomic
molecular vibrations, and it constitutes a convenient model for other physical situations.~

Coulomb plus ring-shaped potential was studied for the three-dimensional motion of a
charged relativistic quantum particle in a noncentral potential [34]. This investigation
is based on a finite-difference version of relativistic quantum mechanics (see references
in [5]) and is a generalization of the results of [16] to the relativistic case.
We emphasize that the Coulomb plus ring-shaped-like potentials \eqref{GrindEQ__1_2_}
are widely used in various fields of physics: in nuclear and particle physics, in atomic
and molecular physics, in condensed matter and chemical physics in the nonrelativistic, as
well as in the relativistic regions.

Possible application of the combined potentials can be the study of the
interaction between deformed pair of the nucleus and spin-orbit coupling
in nuclear physics. Another possibility in the hadronic systems may be the
description of vibrations which can shed light on other physical situations.
In conclusion, the study of the wave functions and the energies in the bound
and continuum states of the interacting systems with the linear combination of
Coulomb and ring-shaped potentials are quite impressive, and it can provide a
deeper and genuine appreciation of the physical properties.

The remainder of the present work proceeds in the following order: in Section II, we give
the information about\textbf{ }the relativistic model of the Coulomb plus ring-shaped like
potential. Next, we present the solutions of the radial KFG equation in Section III.
Then in Section IV, we give the dynamical group of symmetry of the present system.
In Section V, we present the solution of the angle-dependent part of the KFG equation.
In Section VI, we obtain the nonrelativistic limit of the eigenfunctions and the energy
spectrum, and some concluding remarks are stated in Section VII.

\section{The Relativistic model of the Coulomb plus ring-shaped like potential}

The KFG equation with $S(r)$ scalar and $V(r)$ vector potentials has the form[1]
\begin{equation} \label{GrindEQ__2_1_}
\left\{-\hbar ^{2} c^{2} \nabla _{}^{2} +\left[Mc^{2} +S(r)\right]\, ^{2} -\left[E-V(r)\right]\, ^{2} \right\}\; \psi (r)=0.
\end{equation}
In the nonrelativistic limit, this equation transforms into the Schr\"{o}dinger equation for the sum of potentials $V_{N} $ and $S_{N} $, i.e.,
\begin{equation} \label{GrindEQ__2_2_}
\left[-\frac{\hbar ^{2} }{2M} \Delta +V_{N} (r)+S_{N} (r)-E_{N} \right]\; \psi _{N} (r)=0,
\end{equation}
where $E_{N} =\mathop{\lim }\limits_{c\to \infty } (E-Mc^{2} )$, $V_{N} (r)=\mathop{\lim }\limits_{c\to \infty } V(r)$ and $S_{N} (r)=\mathop{\lim }\limits_{c\to \infty } S(r)$.

We present here the exact solutions of the KFG equation with equal scalar and vector
potentials for the sum Coulomb and ring-shaped potentials of the type \eqref{GrindEQ__1_1_}.
We also use an approach based on the Lie algebra of the group $SU(1,1)$, well known to be the
dynamical group for several quantum systems.[35-38] We introduce a tilting transformation that
relates between the physical states and the group sates which constitute a basis of the relevant unitary irreducible representation of $SU(1,1)$.

As done in the literatures [28-33], here we assume that
\begin{equation} \label{GrindEQ__2_3_}
S(r)=V(r)=-\frac{\alpha _{E} }{r} +\frac{f(\theta ,\varphi )}{r^{2} } .
\end{equation}
In terms of dimensionless variables $\rho =r/{\mathchar'26\mkern-10mu\lambda} $, $f_{0} =Mf(\theta ,\varphi )/\hbar ^{2} $ and parameters $\alpha _{0} =\alpha _{E} /\hbar c$, $\varepsilon =E/Mc^{2} $ equation \eqref{GrindEQ__2_1_} takes the form
\begin{equation} \label{GrindEQ__2_4_}
H\psi \equiv \left[-\nabla _{\rho }^{2} +2(1+\varepsilon )\left(-\frac{\alpha _{0} }{\rho } +\frac{f_{0} }{\rho ^{2} } \right)+1-\varepsilon ^{2} \right]\psi (\rho ,\theta ,\varphi )=0.
\end{equation}
Here, the operator $\nabla _{\rho }^{2} $  has the usual form

\[\nabla _{\rho }^{2} =\partial _{\rho }^{2} +\frac{2}{\rho } \partial _{\rho } -\frac{\hat{L}^{2} }{\rho ^{2} },  \partial _{\rho } =\frac{\partial }{\partial \rho } ,   \rho =(\rho ,\theta ,\varphi ),\]
where  $\hat{L}^{2} $ is the square of the angular momentum operator, and it is defined as
\begin{equation} \label{GrindEQ__2_5_}
\hat{L}^{2} =-\Delta _{\theta ,\varphi } =-\, \left[\frac{1}{\sin \theta } \partial _{\theta } (\sin \theta \, \partial _{\theta } )+\frac{1}{\sin ^{2} \theta } \partial _{\varphi }^{2} \right].
\end{equation}
Equation \eqref{GrindEQ__2_4_} does not differ in form from the Schr\"{o}dinger equation \eqref{GrindEQ__2_2_}. It allows separation of variables in the spherical coordinates $\rho ,\theta ,\varphi $ $(0\le \rho <\infty ,\, \, 0\le \theta \le \pi ,\, \, 0\le \varphi <2\pi )$.

Searching the wave function in the form
\begin{equation} \label{GrindEQ__2_6_}
\psi (\rho ,\theta ,\varphi )=\rho ^{-1} R(\rho )F(\theta ,\varphi ),
\end{equation}
one obtains the set of separated differential equations
\begin{equation} \label{GrindEQ__2_7_}
\left(\partial _{\rho }^{2} +\frac{2(1+\varepsilon )\, \alpha _{0} }{\rho } -\frac{g}{\rho ^{2} } -1+\varepsilon ^{2} \right)R=0,
\end{equation}
\begin{equation} \label{GrindEQ__2_8_}
\left(\hat{A}-g\right){\kern 1pt} {\kern 1pt} F(\theta ,\varphi )=0,
\end{equation}
here  $g$ is a separation constant and
\begin{equation} \label{GrindEQ__2_9_}
\hat{A}=-\Delta _{\theta ,\varphi } +M_{0} f(\theta ,\varphi ),  M_{0} =2M(1+\varepsilon )/\hbar ^{2} .
\end{equation}
The operator $\hat{A}$ depends on energy, so its eigenvalues $g$ will also depend on energy, $g=g(\varepsilon ).$ These dependences lead to the fact that an equation determining the energy levels (see formula (5.12)) will be very complicated. However, in the nonrelativistic limit, these dependences disappear, i.e.
\begin{equation} \label{GrindEQ__2_10_}
\mathop{\lim }\limits_{c\to \infty } \, \hat{A}=\hat{A}_{N} =-\Delta _{\theta ,\varphi } +4M\, f_{N} (\theta ,\varphi )/\hbar ^{2}  , \mathop{\lim }\limits_{c\to \infty } g(\varepsilon )=g_{N} .
\end{equation}
The operator $\hat{A}$ commute with the Hamiltonian $H$, i.e. $[H,\hat{A}]=0$. It is responsible for separability of $H$ in the spherical coordinates.

\section{The solutions of the radial Klein-Fock-Gordon equation}

Let us go to the radial wave equation \eqref{GrindEQ__2_7_}. Here we suppose that the $\varepsilon $and $g$ are arbitrary parameters. Putting
\begin{equation} \label{GrindEQ__3_1_}
R(\rho )=\rho ^{\nu } e^{-\sqrt{1-\varepsilon ^{2} } \rho } \, \Omega (\rho ),\; \quad \nu =\frac{1}{2} +\sqrt{\frac{1}{4} +g}
\end{equation}
in \eqref{GrindEQ__2_7_}, there we receive the following equation for the function $\Omega (\rho )$:
\begin{equation} \label{GrindEQ__3_2_}
\rho \, \Omega ''+2(\nu -\sqrt{1-\varepsilon ^{2} } \rho )\, \Omega '+(2(1+\varepsilon )\, \alpha _{0} -2\sqrt{1-\varepsilon ^{2} } \nu )\Omega =0.
\end{equation}
If we comparing this equation with the following equation
\begin{equation} \label{GrindEQ__3_3_}
z\, u''+(\gamma -z)u'-\alpha \, u=0,
\end{equation}
for a confluent hypergeometric function $u=F(\alpha ;\gamma ;z)$, its solutions are expressed in the form:

\[\Omega (\rho )=F(\nu -\frac{(1+\varepsilon )\, \alpha _{0} }{\sqrt{1-\varepsilon ^{2} } } ,\, \, 2\nu \, ;\, \, 2\sqrt{1-\varepsilon ^{2} } \rho ).\]
Consequently,
\begin{equation} \label{GrindEQ__3_4_}
R(\rho )=\rho ^{\nu } e^{-\sqrt{1-\varepsilon ^{2} } \rho } F(\nu -\sqrt{\frac{\varepsilon +1}{\varepsilon -1} } \alpha _{0} ,2\nu \, ;2\sqrt{1-\varepsilon ^{2} } \rho ).
\end{equation}
To analyze the obtained solutions, we consider separately the cases when $\left|\varepsilon \right|<1$ (or $\left|E\right|<Mc^{2} $) and $\left|\varepsilon \right|>1$ (or $\left|E\right|>Mc^{2} $).

1) Let be $\left|\varepsilon \right|<1$. From the condition that the wave function is finite at zero $R(0)=0$ follows that $\nu >0$. This case corresponds to the discrete energy spectrum of our system. Let us find the energy spectrum and wave function. Demand $R(\infty )=0$ for the wave function \eqref{GrindEQ__3_4_} leads to the following condition for energy quantization
\begin{equation} \label{GrindEQ__3_5_}
\nu -\sqrt{\frac{\varepsilon +1}{\varepsilon -1} } \alpha _{0} =-n,\quad \, n=0,1,2,...
\end{equation}
where $n$ denotes the radial quantum number.

It follows from \eqref{GrindEQ__3_5_} the discrete energy levels equation for our system in the case of $0<\left|\varepsilon \right|<1$
\begin{equation} \label{GrindEQ__3_6_}
(n+\nu )\sqrt{1-\varepsilon } =\alpha _{0} \sqrt{1+\varepsilon } ,\quad \; n=0,1,2,...
\end{equation}
where the parameter $\nu $ may only take special values to be determined from \eqref{GrindEQ__2_8_}. We emphasize that for hydrogen-like atoms $\alpha _{0} $ must be replaced by $Z\alpha _{0} $. Then for the energies $-1<\varepsilon <0$ equation \eqref{GrindEQ__3_6_} will be satisfied for sufficiently large values $Z$. For example, when $n=0,\, \; \nu =1$ we get $Z=\frac{\nu }{\alpha _{0} } \sqrt{\frac{1+\left|\varepsilon \right|}{1-\left|\varepsilon \right|} } $. From here when $\left|\varepsilon \right|=0.2$ we find $Z\ge 167$.

Thus, it follows from \eqref{GrindEQ__3_4_} and \eqref{GrindEQ__3_5_} that the radial wave functions corresponding to discrete energy levels \eqref{GrindEQ__3_6_} will have the form
\begin{equation} \label{GrindEQ__3_7_}
R_{n} (\rho )=\tilde{C}_{n} \, \rho ^{\nu } e^{-\sqrt{1-\varepsilon _{n}^{2} } \rho } F(-n,\, 2\nu ;\, 2\sqrt{1-\varepsilon _{n}^{2} } \rho ),
\end{equation}
here\textbf{ }$\varepsilon _{n} $ are roots of equation \eqref{GrindEQ__3_6_}. Using formulas [39]:
\begin{equation} \label{GrindEQ__3_8_}
L_{n}^{\alpha } (x)=\frac{(\alpha +1)_{n} }{n!} F(-n,\alpha +1;x)
\end{equation}
the functions  \textbf{$R_{n} (\rho )$ }can\textbf{ }be expressed by the associated Laguerre polynomials. As a result, we conclude that the normalized radial wave functions, corresponding to the discrete energy spectrum \eqref{GrindEQ__3_6_} are
\begin{equation} \label{GrindEQ__3_9_}
R_{n} (\rho )=C_{n} \rho ^{\nu } e^{-\sqrt{1-\varepsilon _{n}^{2} } \rho } L_{n}^{2\nu -1} (2\sqrt{1-\varepsilon _{n}^{2} } \rho ),\; \quad
\end{equation}
where $C_{n} $ is the normalization constant
\begin{equation} \label{GrindEQ__3_10_}
\quad C_{n} =2^{\nu } \sqrt{\frac{n\, !\, \, (1-\varepsilon _{n}^{2} )^{\nu +1/2} }{2(n+\nu )\Gamma (n+2\nu )} } .
\end{equation}
It is found from the following condition
\begin{equation} \label{GrindEQ__3_11_}
\int _{0}^{\infty }R_{n}^{2} (\rho )d\rho =1 .
\end{equation}
To calculate this integral, the standard trick was used, i.e., recurrence relation
\begin{equation} \label{GrindEQ__3_12_}
xL_{n}^{\alpha } (x)=(2n+\alpha +1)L_{n}^{\alpha } (x)-(n+1)L_{n+1}^{\alpha } (x)-(n+1)L_{n-1}^{\alpha } (x)
\end{equation}
and orthogonality properties for associated Laguerre polynomials:

\noindent
\begin{equation} \label{GrindEQ__3_13_}
\int _{0}^{\infty }x^{\alpha } e^{-x} L_{n}^{\alpha } (x)L_{m}^{\alpha } dx =\frac{\Gamma (n+\alpha +1)}{n!} \delta _{nm} .
\end{equation}
2) Let now $\left|\varepsilon \right|>1$. In this case, the energy spectrum will be continuous, and the corresponding wave functions are obtained from expression \eqref{GrindEQ__3_4_}. For example, for $\varepsilon >1$ (in this case $\nu $  is a real parameter)
\begin{equation} \label{GrindEQ__3_14_}
R(\rho )=C\rho ^{\nu } e^{-i\tau \rho } F(\nu +i(1+\varepsilon )\, \alpha _{0} /\sqrt{\varepsilon ^{2} -1} ,\; 2\nu \, ;\; 2i\sqrt{\varepsilon ^{2} -1} \cdot \rho ), \quad \tau =\sqrt{\varepsilon ^{2} -1} .
\end{equation}
Radial wave functions corresponding to energy values $\varepsilon =0,\, \pm 1$, can be obtained from \eqref{GrindEQ__3_4_} by using the corresponding passage to the limit. For example, for the value $\varepsilon =1$ we find
\begin{equation} \label{GrindEQ__3_15_}
\mathop{\lim }\limits_{\varepsilon \to 1} R(\rho )=C'\rho ^{-\nu _{1} +1} J_{2\nu _{1} -1} (4\sqrt{\alpha {}_{0}^{} \, \rho } ),
\end{equation}
where $\nu _{1} =\mathop{\lim }\limits_{\varepsilon \to 1} \nu $, a  $J_{\nu } (z)$ are well known Bessel functions.
In the derivation of \eqref{GrindEQ__3_15_}, we used the easily proved limit formula
\begin{equation} \label{GrindEQ__3_16_}
\mathop{\lim }\limits_{\alpha \to 0} F(\nu -\frac{2(1+\varepsilon )\, \alpha _{0} }{\alpha } ,\, \gamma \, \, ;\alpha \, x)=\Gamma (\gamma +1)(a\, x)^{-\gamma +1} J_{\gamma -1} (2\sqrt{a\, x}). \eqref{GrindEQ__3_16_}
\end{equation}
The asymptotic behavior of Bessel functions at zero and at infinity are given by the formulas, respectively [39]

\[\mathop{J_{\nu } (z)}\limits_{z\to 0} \approx (z/2)^{\nu } ,\]
\begin{equation} \label{GrindEQ__3_17_}
\mathop{J_{\nu } (z)}\limits_{z\to \infty } \approx \sqrt{\frac{2}{\pi z} } \left[\cos (z-\frac{\nu \pi }{2} -\frac{\pi }{4} )-\frac{1}{z} (\nu ^{2} -1/4)\sin (z-\frac{\nu \pi }{4} -\frac{\pi }{4} )\right].
\end{equation}
It follows that the wave function \eqref{GrindEQ__3_16_} is nonquadratic integrable at zero and
at infinity. This means that the radial equation \eqref{GrindEQ__2_7_} does not have quadratically
integrable solutions, therefore, a charged particle (for example, electron) in the Coulomb plus
ring-shaped potential field \eqref{GrindEQ__2_3_} of an arbitrary charge $\alpha _{E} =Ze^{2} $
does not have bound states with energy$E=Mc^{2} $, i.e., with zero binding energy. The nature of
this phenomenon, as in the case of the Dirac equation considered in [40], is associated with the
long-range nature of the Coulomb potential. The fact is that this potential generates an infinite
set of bound states with energy equations defined by formula \eqref{GrindEQ__3_6_}. This set accumulates at a point $E=Mc^{2} $,
but does not reach this point. Such a picture remains valid for any value of the charge $Z$.
This conclusion can also be obtained based on the wave function \eqref{GrindEQ__3_14_}.

\section{Dynamical Symmetry Group}

Let us now consider the radial equation \eqref{GrindEQ__2_7_} by help of $SU(1,1)$ Lie algebra. The generators of $SU(1,1)$ algebra
may be realized as [35]
\begin{equation} \label{GrindEQ__4_1_}
\begin{array}{l} {K_{0} \equiv \Gamma _{0} =\frac{1}{2} \left(-\rho \, \partial _{\rho }^{2} +\frac{g}{\rho } +\rho \right),} \\ {K_{1} \equiv \Gamma _{4} =\frac{1}{2} \left(-\rho \, \partial _{\rho }^{2} +\frac{g}{\rho } -\rho \right),} \\ {K_{2} \equiv T=-i\rho \, \partial _{\rho } .} \end{array}
\end{equation}
By a direct check, one can verify that these operators satisfy the commutation relations
\begin{equation} \label{GrindEQ__4_2_}
\left[\Gamma _{0} ,\Gamma _{4} \right]=iT,\quad \left[T,\Gamma _{0} \right]=i\Gamma _{4} ,\; \; \left[\Gamma _{4} ,T\right]=-i\Gamma _{0} .
\end{equation}
The Casimir operator[36] is
\begin{equation} \label{GrindEQ__4_3_}
C_{2} =\Gamma _{0}^{2} -\Gamma _{4}^{2} -T^{2} =s\, (s-1).
\end{equation}
We denote the states of a positive discrete series as $\left|n,s>\right. $ such that
\begin{equation} \label{GrindEQ__4_4_}
\begin{array}{l} {\Gamma _{0} \left|n,s>=(n+s)\right. \left|n,s>\right. ,} \\ {C_{2} \left|n,s>=s(s-1)\right. \left|n,s>,\right. } \end{array}
\end{equation}
where $s$ is the Bargmann index, $s>0$ and $n=0,1,2,...$. It should be note that in our case from Eq. \eqref{GrindEQ__4_1_}, we obtain  $C_{2} =g=\nu \, (\nu -1)$, so $s=\nu $. The equation \eqref{GrindEQ__2_7_} can be written with the help of generators \eqref{GrindEQ__3_1_} as
\begin{equation} \label{GrindEQ__4_5_}
\left[(2-\varepsilon ^{2} )\Gamma _{0} +\varepsilon ^{2} \Gamma _{4} -2(1+\varepsilon )\, \alpha _{0} \right]R=0.
\end{equation}
Let us introduce a supplementary parameter $\theta $ and perform a tilting transformation
\begin{equation} \label{GrindEQ__4_6_}
\tilde{R}=SR,\, \; S=e^{-i\theta {\kern 1pt} T} .
\end{equation}
From the commutation relations \eqref{GrindEQ__4_2_} and the formula
\begin{equation} \label{GrindEQ__4_7_}
e^{\hat{A}} \hat{B}\; e^{-\hat{A}} =\hat{B}+[\hat{A},\hat{B}]+\frac{1}{2!} [\hat{A},[\hat{A},\hat{B}]]+...,
\end{equation}
where $\hat{A}$ and $\hat{B}$ are any two operators, it follows that
\begin{equation}\label{GrindEQ__4_8_}
\begin{array}{l} {e^{-i\theta \, {\kern 1pt} T} \Gamma _{0} {\kern 1pt} e^{i\theta {\kern 1pt} \, T} =\Gamma _{0} \cosh \theta +\Gamma _{4} \sinh \theta ,} \\ {e^{-i\theta \, {\kern 1pt} T} \Gamma _{4} {\kern 1pt} e^{i\theta {\kern 1pt} \, T} =\Gamma _{4} \cosh \theta +\Gamma _{0} \sinh \theta ,} \\ {e^{-i\beta \; \Gamma _{0} } Te^{i\beta \; \Gamma _{0} } =T\cos \beta -\Gamma _{4} \sin \beta ,}  \\ 
{e^{-i\beta \; \Gamma _{0} } \Gamma _{4} e^{i\beta \; \Gamma _{0} } =\Gamma _{4} \cos \beta +T\sin \beta ,} \\ {e^{-i\alpha \; \Gamma _{4} } \Gamma _{0} e^{i\alpha \; \Gamma _{4} } =\Gamma _{0} \cosh \alpha -T\sinh \alpha ,} \\ {e^{-i\alpha \; \Gamma _{4} } Te^{i\alpha \; \Gamma _{4} } =T\cosh \alpha -\Gamma _{0} \sinh \alpha .} \end{array}
\end{equation}
%
Through using the formula \eqref{GrindEQ__4_8_}, it is easily to verified that the equation \eqref{GrindEQ__4_5_} becomes
\begin{equation} \label{GrindEQ__4_9_}
\left[((2-\varepsilon ^{2} )\cosh \theta +\varepsilon ^{2} \sinh \theta )\Gamma _{0} +((2-\varepsilon ^{2} )\sinh \theta +\varepsilon ^{2} \cosh \theta )\Gamma _{4} -2(1+\varepsilon )\, \alpha _{0} \right]{\kern 1pt} {\kern 1pt} \tilde{R}=0.
\end{equation}
To solve equation \eqref{GrindEQ__4_9_} in an algebraic way, we consider separately the cases when $\left|\varepsilon \right|<1$ (or $\left|E\right|<Mc^{2} $) and $\left|\varepsilon \right|>1$ (or $\left|E\right|>Mc^{2} $).

1) When $\left|\varepsilon \right|<1$ (discrete spectrum) in \eqref{GrindEQ__4_9_}, a compact generator $\Gamma _{0} $ can be diagonalized. Setting the coefficient of $\Gamma _{4} $ to zero we obtain  $\tanh \theta =-\varepsilon ^{2} /(2-\varepsilon ^{2} )$ or $e^{\theta } =\sqrt{1-\varepsilon ^{2} } $. As a result, we have
\begin{equation} \label{GrindEQ__4_10_}
\left(\sqrt{1-\varepsilon ^{2} } \Gamma _{0} -(1+\varepsilon )\, \alpha _{0} \right){\kern 1pt} {\kern 1pt} \tilde{R}=0.
\end{equation}
It follows from \eqref{GrindEQ__4_10_} the energy levels equation \eqref{GrindEQ__3_6_}.

2) When $\left|\varepsilon \right|>1$ non-compact generator $\Gamma _{4} $ is
diagonalized, having a continuous real spectrum $\lambda \in R$. In this case, we equate
to zero the coefficient of the operator $\Gamma _{0} $, i.e. $(2-\varepsilon ^{2} )\cosh \theta +\varepsilon ^{2} \sinh \theta =0$.
From here we get $\tanh \theta =-(2-\varepsilon ^{2} )/\varepsilon ^{2} $ and $e^{\theta } =\sqrt{\varepsilon ^{2}-1 } $.
Moreover, equation \eqref{GrindEQ__4_9_} takes the form
\begin{equation} \label{GrindEQ__4_11_}
(\sqrt{(\varepsilon ^{2}-1 )} \Gamma _{4} -(1+\varepsilon )\, \alpha _{0} )\tilde{R}_{1} =0.
\end{equation}
Consequently,
\begin{equation} \label{GrindEQ__4_12_}
E_{\lambda } =\frac{\lambda ^{2} +\alpha _{0}^{2} }{\lambda ^{2} -\alpha _{0}^{2} } Mc^{2} .
\end{equation}
We emphasize that the generators $\Gamma _{0} $, $\Gamma _{4} $ and $T$ have no nonrelativistic limit: they diverge at $c\to \infty $. Generators $\Gamma '_{0} $, $\Gamma '_{4} $ and $T'$ , which have the correct nonrelativistic limit can be obtained from $\Gamma _{0} $ , $\Gamma _{4} $ and  $T$ by help of unitary transformation  (compare with [41]). Unitary operator $U$, performing such a transformation has the form
\begin{equation} \label{GrindEQ__4_13_}
U=e^{-i\omega \, T} , \tanh \omega =\frac{1-\varepsilon ^{2} -\alpha _{0}^{2} }{1-\varepsilon ^{2} +\alpha _{0}^{2} } , e^{\omega } =\sqrt{1-\varepsilon ^{2} } /\alpha _{0} .
\end{equation}
Then, we will have
\[\Gamma '_{0} =U\, \Gamma _{0} U^{-1} =\Gamma _{0} \cosh \omega +\Gamma _{4} \sinh \omega =\frac{1}{2} (-\xi \, \partial _{\xi }^{2} +\frac{g}{\xi } +\xi ),\]
\begin{equation} \label{GrindEQ__4_14_}
\Gamma '_{4} =U\, \Gamma _{4} U^{-1} =\Gamma _{4} \cosh \omega +\Gamma _{0} \sinh \omega =\frac{1}{2} (-\xi \, \partial _{\xi }^{2} +
\frac{g}{\xi } -\xi),
\end{equation}
\[T'=UTU^{-1} =T=-i\xi \, \partial _{\xi } .\]
In the formulas \eqref{GrindEQ__4_14_} the following dimensionless variable is introduced $\xi ={r \mathord{\left/{\vphantom{r a_{0} }}\right.\kern-\nulldelimiterspace} a_{0} } $, where $a_{0} ={\hbar ^{2}  \mathord{\left/{\vphantom{\hbar ^{2}  M\alpha _{e} }}\right.\kern-\nulldelimiterspace} M\alpha _{e} } $ is a Bohr radius. It is associated with $\rho =r/{\mathchar'26\mkern-10mu\lambda} $ in the following way $\xi =\alpha _{0} \rho $. We can also write relations \eqref{GrindEQ__4_14_} in matrix form
\begin{equation} \label{GrindEQ__4_15_}
\left(\begin{array}{c} {\Gamma '_{0} } \\ {\Gamma '_{4} } \\ {T'} \end{array}\right)=\left(\begin{array}{ccc} {\cosh \omega } & {\sinh \omega } & {0} \\ {\sinh \omega } & {\cosh \omega } & {0} \\ {0} & {0} & {1} \end{array}\right)\, \left(\begin{array}{c} {\Gamma _{0} } \\ {\Gamma _{4} } \\ {T} \end{array}\right).
\end{equation}
As is known, under unitary transformations, the Casimir operator remains unchanged, i.e. $C'_{2} =UC_{2} U^{-1} =C_{2} =g$.

Equation \eqref{GrindEQ__2_7_} can also be solved algebraically using the generators $\Gamma '_{0} $, $\Gamma '_{4} $ and $T'$. To do this, we rewrite it in the form
\begin{equation} \label{GrindEQ__4_16_}
\left(\alpha _{1} \Gamma '_{0} +\alpha _{2} \Gamma '_{4} -2(1+\varepsilon )\, \alpha _{0} \right)R=0,     \alpha _{1,2} =\alpha _{0} \pm (1-\varepsilon ^{2} )/\alpha _{0} .
\end{equation}
We now perform the tilting transformation to remove the noncompact generator $\Gamma _{4} $ [35-38]. To this end, we define
\begin{equation} \label{GrindEQ__4_17_}
\tilde{R'}=S'R,     S'=e^{-i\theta '\, T} .
\end{equation}
In view of formulas \eqref{GrindEQ__4_8_}, we rewrite equation \eqref{GrindEQ__4_16_} in the form
\begin{equation} \label{GrindEQ__4_18_}
[(\alpha _{1} \cosh \theta '+\alpha _{2} \sinh \theta ')\Gamma '_{0} +(\alpha _{1} \sinh \theta '+\alpha _{2} \cosh \theta ')\Gamma '_{4} -2(1+\varepsilon )\, \alpha _{0} ]\tilde{R'}=0.
\end{equation}
If we choose now $\theta '$ as
\begin{equation} \label{GrindEQ__4_19_}
\tanh \theta '=-\frac{\alpha _{2} }{\alpha _{1} } =\frac{1-\varepsilon ^{2} -\alpha _{0}^{2} }{1-\varepsilon ^{2} +\alpha _{0}^{2} } ,  e^{\theta '} =\sqrt{1-\varepsilon ^{2} } /\alpha _{0} ,
\end{equation}
which corresponds to the case $0<\left|\varepsilon \right|<1$, then in \eqref{GrindEQ__4_18_} the generator $\Gamma '_{0} $ is diagonalized, i.e.
\begin{equation} \label{GrindEQ__4_20_}
(\sqrt{1-\varepsilon ^{2} } \Gamma '_{0} -(1+\varepsilon )\, \alpha _{0} )\tilde{R'}=0.
\end{equation}
We obtain from \eqref{GrindEQ__4_20_} again the discrete energy spectrum \eqref{GrindEQ__3_6_} pure algebraically.

\noindent Now we give the relationship between the functions $\tilde{R}$ and $\tilde{R'}$. It has the form
\begin{equation} \label{GrindEQ__4_21_}
\tilde{R'}=\tilde{S}\tilde{R}, \tilde{S}=S(S')^{-1} =e^{-i\tilde{\theta }T} , \tilde{\theta }=\theta -\theta ',
\end{equation}
moreover $\tanh \tilde{\theta }={(\alpha _{0}^{2} -1) \mathord{\left/{\vphantom{(\alpha _{0}^{2} -1) (\alpha _{0}^{2} +1)}}\right.\kern-\nulldelimiterspace} (\alpha _{0}^{2} +1)} $ and  $e^{\tilde{\theta }} =\alpha _{0} $. Consequently,
\begin{equation} \label{GrindEQ__4_22_}
\tilde{R'}_{n} (\xi )=e^{-i\tilde{\theta }T} \tilde{R}_{n} (\rho )=\tilde{R}_{n} (\alpha _{0} \rho ).
\end{equation}
When $\left|\varepsilon \right|>1$, we can diagonalized generator $\Gamma '_{4} $, and we get the equation
\begin{equation} \label{GrindEQ__4_23_}
(\sqrt{\varepsilon ^{2} -1} \Gamma '_{4} -(1+\varepsilon )\, \alpha _{0} )\tilde{R'}=0.
\end{equation}
From here follows a continuous energy spectrum \eqref{GrindEQ__4_12_}.

\section{The solutions of the angular equation}

We now investigate the solutions of the angle-dependent equation \eqref{GrindEQ__2_8_}. For this aim we choose a function $f(\theta ,\varphi )$
in the form
\begin{equation} \label{GrindEQ__5_1_}
f(\theta ,\varphi )=\frac{\gamma +\beta \cos \theta +\alpha \cos ^{2} \theta }{\sin ^{2} \theta } .
\end{equation}
In this case $z$- projection of the angular momentum operator $\hat{L}_{z} =-i\partial _{\varphi } $ will also commute with the Hamiltonian $H$. Thus, the variables in equation \eqref{GrindEQ__2_8_} can be further separated in the usual way
\begin{equation} \label{GrindEQ__5_2_}
F(\theta ,\varphi )=\Theta (\theta )\, e^{im\varphi } /\sqrt{2\pi } .
\end{equation}
Here $m$\textit{ }is the usual magnetic quantum number and is integer. In terms of the new variable $x=\cos \theta $, we obtain the following equation for the function $\Theta (\theta )\equiv \Theta (x)$:
\begin{equation} \label{GrindEQ__5_3_}
[\partial _{x}^{2} +\frac{\tilde{\tau }(x)}{\sigma (x)} \partial _{x} +\frac{\tilde{\sigma }(x)}{\sigma ^{2} (x)} ]\, \Theta (x)=0,
\end{equation}
where $\sigma =1-x^{2} $, $\tilde{\tau }=-2x$, $\tilde{\sigma }=-c_{2} x^{2} -c_{1} x+c_{0} $,
\begin{equation} \label{GrindEQ__5_4_}
c_{2} =g+M_{0} \alpha , c_{1} =M_{0} \beta ,  c_{0} =g-m^{2} -M_{0} \gamma .
\end{equation}
The orthonormalized solutions of equation \eqref{GrindEQ__5_3_} were found in [34]. We give their explicit form
\begin{equation} \label{GrindEQ__5_5_}
\Theta _{km} (\theta )=c_{km} (\sin ^{2} \frac{\theta }{2} )^{-A_{1} } (\cos ^{2} \frac{\theta }{2} )^{-A_{2} } P_{k}^{(-2A_{1} ,-2A_{2} )} (\cos \theta ), k=0,1,2,...
\end{equation}
Parameters $A_{1,2} <0$ are defined by expressions
\begin{equation} \label{GrindEQ__5_6_}
A_{1,2} =-\frac{1}{2} \sqrt{m^{2} +M_{0} (\alpha +\gamma \pm \beta )} .
\end{equation}
Because the $M_{0} =2M(1+\varepsilon )/\hbar ^{2} $, then these parameters, in contrary to the results of [34], are depend on energy. A function $P_{k}^{(\alpha ,\beta )} (x)$, which are the Jacobi polynomials, satisfies the following orthogonality condition.[42]
\begin{equation} \label{GrindEQ__5_7_}
\int _{-1}^{1}(1-x)^{\alpha } (1+x)^{\beta } P_{k}^{(\alpha ,\beta )} (x) P_{k'}^{(\alpha ,\beta )} (x)dx=d_{k}^{2} \delta _{kk'} ,
\end{equation}
where the square of the norm of the polynomials is
\begin{equation} \label{GrindEQ__5_8_}
d_{k}^{2} =\frac{2^{\alpha +\beta +1} \Gamma (k+\alpha +1)\Gamma (k+\beta +1)}{(2k+\alpha +\beta +1)k!\Gamma (k+\alpha +\beta +1)} ,
\alpha >-1, \beta >-1.
\end{equation}
It follows from condition \eqref{GrindEQ__5_7_} that the angular wave functions \eqref{GrindEQ__5_5_} for distinct values of  $k$  satisfy orthonormal condition
\begin{equation} \label{GrindEQ__5_9_}
\int _{-1}^{1}\Theta _{km} (x)\Theta _{k'm} (x)dx=\delta _{kk'}  .
\end{equation}
Here the normalization constant $c_{km} $ is equal to
\begin{equation} \label{GrindEQ__5_10_}
c_{km} =\sqrt{\frac{(k-A_{1} -A_{2} +1/2)k!\Gamma (k-2A_{1} -2A_{2} +1)}{\Gamma (k-2A_{1} +1)\Gamma (k-2A_{2} +1)} } .
\end{equation}
As shown in the Ref. [34], separation constant $g$ depends on $k$ and $m$  in this form
\begin{equation} \label{GrindEQ__5_11_}
\begin{array}{l} {g\equiv g_{km} (\varepsilon )=k[k+1+\sqrt{m^{2} +M_{0} (\alpha +\gamma +\beta )} +\sqrt{m^{2} +M_{0} (\alpha +\gamma -\beta )} ]+} \\ {+\frac{1}{2} [\sqrt{m^{2} +M_{0} (\alpha +\gamma +\beta )} +\sqrt{m^{2} +M_{0} (\alpha +\gamma -\beta )} ]+} \\ {+\frac{1}{2} \sqrt{[m^{2} +M_{0} (\alpha +\gamma +\beta )][m^{2} +M_{0} (\alpha +\gamma -\beta )]} +\frac{1}{2} [m^{2} +M_{0} (\gamma -\alpha )].} \end{array}
\end{equation}
Thus, the exact discrete energy eigenvalues of the KFG equation for our system are defined as follows
\begin{equation} \label{GrindEQ__5_12_}
(n+\frac{1}{2} +\sqrt{\frac{1}{4} +g_{km} (\varepsilon )} )\sqrt{1-\varepsilon } =\alpha _{0} \sqrt{1+\varepsilon } .
\end{equation}
This equation implies that the calculation of the energy levels becomes very complicated. We consider two special cases of formulas \eqref{GrindEQ__3_6_} and \eqref{GrindEQ__5_5_}.

\noindent \textbf{5.1.}
For $f(\theta ,\varphi )=0$ in equation \eqref{GrindEQ__2_3_}, the potential we used becomes Coulomb potential. In this case $g=l(l+1)$, $\nu =l+1$ and we will have $A_{1} =A_{2} =-\left|m\right|/2$, $l=k+\left|m\right|$,
\begin{equation} \label{GrindEQ__5_13_}
\Theta _{lm} (\theta )=(-1)^{\frac{m-\left|m\right|}{2} } \frac{1}{2^{\left|m\right|} l!} \sqrt{\frac{2l+1}{2} (l-\left|m\right|)!(l+\left|m\right|)!\, } (\sin \theta )^{\left|m\right|} P_{l-\left|m\right|}^{(\left|m\right|,\left|m\right|)} (\cos \theta ),
\end{equation}
Letting $\alpha =\beta =\gamma =0$ in \eqref{GrindEQ__5_12_} and solving the corresponding energy equation, we obtain for the Coulomb potential
\begin{equation} \label{GrindEQ__5_14_}
\frac{E_{nl}}{Mc^2} = \varepsilon_{nl}=\left[1-\frac{2\alpha _{0}^{2} }{\alpha _{0}^{2} +\hbar ^{2} c^{2} (n+l+1)^{2} } \right].
\end{equation}
At the nonrelativistic limit, we get the following formula
\begin{equation} \label{GrindEQ__5_15_}
\mathop{\lim }\limits_{c\to \infty } (E_{n} -Mc^{2} )=E_{Nn} =-\frac{2M\alpha _{E}^{2} }{\hbar ^{2} (n+l+1)^{2} } .
\end{equation}

\noindent
The corresponding radial wave functions have the form
\begin{equation} \label{GrindEQ__5_16_}
R_{nl} (\rho )=\sqrt{\frac{n!(1-\varepsilon _{n}^{2} )^{l+3/2} }{(n+l+1)\Gamma (n+2l+2)} } (2\rho )^{l+1} e^{-\sqrt{1-\varepsilon _{n}^{2} } \rho } L_{n}^{2l+1} (2\sqrt{1-\varepsilon _{n}^{2} } \rho ).
\end{equation}
Thanks  to the connecting formula[39] between the Jacobi polynomials $P_{n}^{(\alpha ,\alpha )} (x)$ and the Gegenbauer polynomials $C_{n}^{\lambda } (x)$
\begin{equation} \label{GrindEQ__5_17_}
(\lambda +\frac{1}{2} )_{n} C_{n}^{\lambda } (x)=(2\lambda )_{n} P_{n}^{(\lambda -\, \frac{1}{2} ,\lambda -\, \frac{1}{2} )} (x), (\lambda )_{n} =\frac{\Gamma (n+\lambda )}{\Gamma (\lambda )} ,
\end{equation}

we can express the functions $\Theta _{lm} (\theta )$ \eqref{GrindEQ__5_13_} through the polynomials $C_{n}^{\lambda } (x)$, i.e.
\begin{equation} \label{GrindEQ__5_18_}
\Theta _{lm} (\theta )=(-1)^{\frac{m-\left|m\right|}{2} } 2^{\left|m\right|} \Gamma (\left|m\right|+\frac{1}{2} )\sqrt{\frac{2l+1}{2\pi } \frac{(l-\left|m\right|)!}{(l+\left|m\right|)!} } (\sin \theta )^{\left|m\right|} C_{l-\left|m\right|}^{\left|m\right|+1/2} (\cos \theta ).
\end{equation}
On the other hand, there is the connecting formula [39]
\begin{equation} \label{GrindEQ__5_19_}
P_{l}^{\left|m\right|} (\cos \theta )=\frac{(-2)^{\left|m\right|} }{\sqrt{\pi } } \Gamma (\left|m\right|+1/2)(\sin \theta )^{\left|m\right|} C_{l-\left|m\right|}^{\left|m\right|+1/2} (\cos \theta )
\end{equation}
between the Gegenbauer polynomials $C_{n}^{\lambda } (\cos \theta )$ and the associated Legendre functions $P_{l}^{\left|m\right|} (\cos \theta )$. We have instead of the formula \eqref{GrindEQ__5_18_}[1]
\begin{equation} \label{GrindEQ__5_20_}
\Theta _{lm} (\theta )=(-1)^{\frac{m+\left|m\right|}{2} } \sqrt{\frac{2l+1}{2} \frac{(l-\left|m\right|)!}{(l+\left|m\right|)!} } P_{l}^{\left|m\right|} (\cos \theta ).
\end{equation}
Hence, the total wave functions take the form
\begin{equation} \label{GrindEQ__5_21_}
\begin{array}{l} {\psi _{nkm} (r,\theta ,\varphi )=} \\
{=C_{n} \left(2\sqrt{1-\varepsilon _{n}^{2} } \rho \right)^{v} e^{-\sqrt{1-\varepsilon _{n}^{2} } \rho } L_{n}^{2v-1} (2\sqrt{1-\varepsilon _{n}^{2} } \rho )\left(\sin ^{2} \frac{\theta }{2} \right)^{-A_{1} } \left(\cos ^{2} \frac{\theta }{2} \right)^{-A_{2} } P_{k}^{(-2A_{1} ,-2A_{2} )} \left(\cos \theta \right)\, e^{im\varphi } } \end{array}
\end{equation}
in the case of a discrete spectrum, and
\begin{equation} \label{GrindEQ__5_22_}
\begin{array}{l} {\psi _{km} (r,\theta ,\varphi )=} \\ {=C_{} \rho ^{v} e^{-i\tau \rho } F\left(v+2i\sqrt{\frac{\varepsilon +1}{\varepsilon -1} } \alpha _{0} ,\, 2v;\, 2i\sqrt{\varepsilon ^{2} -1} \rho \right)\left(\sin ^{2} \frac{\theta }{2} \right)^{-A_{1} } \left(\cos ^{2} \frac{\theta }{2} \right)^{-A_{2} } P_{k}^{(-2A_{1} ,-2A_{2} )} \left(\cos \theta \right)\, e^{im\varphi } } \end{array}
\end{equation}
in the case of a continuous spectrum.

\noindent\textbf{5.2.}
At $\gamma =\beta =0$ the equation \eqref{GrindEQ__5_5_} coincides with equation (16) of Ref. [32]. (The relationship between our parameter $\nu $ and parameter $L$ of Ref. [32] is as follows $\nu =L+1$.)

\section{The nonrelativistic limit }

Let us now show that the energy spectrum and the radial wave functions considered in previous sections in the nonrelativistic limit $c\to \infty $ pass into the energy spectrum and the radial wave functions of the hydrogen like atoms respectively. We have
\begin{eqnarray}
\mathop{\lim }\limits_{c\to \infty } A_{1,2} =\left. A_{1,2} \right|_{\varepsilon =1} =A_{N1,2} ,  \mathop{\lim }\limits_{c\to \infty } \nu =\left. \nu \right|_{\varepsilon =1} =\nu _{N} ,   \mathop{\lim }\limits_{c\to \infty } 2\sqrt{1-\varepsilon _{n}^{2} } \rho =\frac{2\rho _{N} }{n+\nu _{N} },
\nonumber \\
\rho _{N} =\frac{2M\alpha _{E} }{\hbar ^{2} } r,\mathop{\lim }\limits_{c\to \infty } (E_{n} -Mc^{2} )=E_{Nn} =
-\frac{2M\alpha _{E}^{2} }{\hbar ^{2}(n+\nu_N)^2},  \nonumber
\end{eqnarray}

\begin{equation} \label{GrindEQ__6_1_}
\mathop{\lim }\limits_{c\to \infty } (E_{\lambda } -Mc^{2} )=E_{N\lambda } =\frac{2M\alpha _{E}^{2} }{\hbar ^{2} \lambda ^{2} } ,
\end{equation}
\begin{eqnarray}
\mathop{\lim }\limits_{c\to \infty } \sqrt{\frac{\varepsilon +1}{\varepsilon -1} } \alpha _{0} =\sqrt{\frac{2M\alpha _{E}^{2} }{\hbar ^{2} E_{N\lambda } ^{2} } } \equiv \frac{1}{k} ,  \mathop{\lim }\limits_{c\to \infty } C_{n} =C_{Nn} =\sqrt{\frac{n!}{2(n+\nu _{N} )\Gamma (n+2\nu _{N} )} } ,
 \nonumber  \\
\mathop{\lim }\limits_{c\to \infty } 2i\sqrt{\varepsilon ^{2} -1} \rho =2ik\rho _{N} . \nonumber
\end{eqnarray}
Accounting the Eq. \eqref{GrindEQ__6_1_}, we obtain that nonrelativistic limits of wave functions are equal to
\begin{equation} \label{GrindEQ__6_2_}
\mathop{\mathop{\lim }\limits_{c\to \infty } R_{n} (\rho )=R_{Nn} (\rho _{N} )}\limits_{} =C_{Nn} \left(\frac{2\rho _{N} }{n+\nu _{N} } \right)^{\nu _{N} } e^{-\, \frac{\rho _{N} }{(n+\nu _{N} )} } L_{n}^{2\nu _{N} -\, 1} \left(\frac{2\rho _{N} }{n+\nu _{N} } \right),
\end{equation}
\[\mathop{\mathop{\lim }\limits_{c\to \infty } R(\rho )=R_{N} (\rho _{N} )}\limits_{} =C_{N} (\rho _{N} )^{\nu _{N} } e^{-ik\rho _{N} } F\left(\nu _{N} +\frac{i}{k} ,\, 2\nu _{N} ;\, 2ik\rho _{N} \right).\]
We now give the nonrelativistic limit of operators \eqref{GrindEQ__4_14_}

\[\Gamma _{0}^{N} =\mathop{\lim }\limits_{c\to \infty } \Gamma '_{0} =\frac{1}{2} (-\xi \, \partial _{\xi }^{2} +\frac{g_{N} }{\xi } +\xi ),\]
\begin{equation} \label{GrindEQ__6_3_}
\Gamma _{4}^{N} =\mathop{\lim }\limits_{c\to \infty } \Gamma '_{4} =\frac{1}{2} (-\xi \, \partial _{\xi }^{2} +\frac{g_{N} }{\xi } -\xi ),
\end{equation}
\[T^{N} =\mathop{\lim }\limits_{c\to \infty } T'=T=-i\xi \, \partial _{\xi } .\]
\textbf{}

\section{Conclusion}

In this paper, the exact solutions of the KFG equation are obtained for equal scalar and vector Coulomb plus ring-shaped potentials.
The KFG equation coincides in form with the Schr\"{o}dinger equation for this potential. We solved the wave
equation in the standard way by the separation of variables. For the radial wave equation, we constructed
a dynamical symmetry group $SU(1,1)$, which in turn allowed us to find the corresponding discrete and continuous
 energy spectrum of the system by purely algebraically. We also found an explicit form of the wave functions
 corresponding to the discrete and continuous energy spectra.
They were expressed via associated Laguerre polynomials or by confluent hypergeometric functions.

It is found that the radial wave functions, corresponding to both discrete and continuous energy eigenvalues, have
 the correct nonrelativistic limits. Angular wave functions $\Theta _{km} (\theta )$ are expressed in terms of
 Jacobi polynomials and they are a generalization of the result of [16] to the relativistic case. Therefore, in
the nonrelativistic limit, wave functions $\Theta _{km} (\theta )$ reproduce the results in Ref. [16].

\noindent It was also shown that radial part of the equation of the motion possesses $SU(1,1)$ dynamical symmetry
 groups. In this line, our work may enable to provide a promising avenue in many branches of physics.

\noindent  In the literature there are special cases of our results. For instance, 1) at  $f(\theta ,\varphi )=0$ from
our results one obtains the known results of the corresponding problem (see for example [1]); 2) at $\beta =\gamma =0$ from formulas \eqref{GrindEQ__3_6_} and \eqref{GrindEQ__3_9_} for the discrete energy spectrum and radial wave functions, the corresponding formulas of Ref. [32] are obtained. 3) at $\beta =\alpha =0$ we obtain from our formulas \eqref{GrindEQ__3_6_} and \eqref{GrindEQ__3_7_} the corresponding formulas (11) and (15) of Ref. [43].

The main results of this paper are the explicit and closed form expressions for the energy spectrum and corresponding wave functions. We have also shown the radial part of the equation of the motion which possesses $SU(1,1)$ dynamical symmetry groups.

\noindent

\noindent \textbf{References}

\noindent [1] W. Greiner, Relativistic Quantum Mechanics, 3rd. ed., Springer, Berlin, 2000.

\noindent [2] V. G. Bagrov and D. M. Gitman, Exact Solutions of Relativistic Wave Equations, KluwerAcademic Publishers, Dordrecht, 1990.

\noindent [3] H. Feshbach and F. Villars, Rev. Mod. Phys. \textbf{30} (1958) 24.

\noindent [4] V. G. Kadyshevsky, R. M. Mir-Kasimov, and N. B. Skachkov, Phys. of Elem. Part. and At. Nucl. \textbf{12} (1972) 635.

\noindent [5] E. D. Kagramanov, R. M. Mir-Kasimov, and Sh. M. Nagiyev, J. Math. Phys. \textbf{31} (1990) 1733.

\noindent [6] Sh. M. Nagiyev, J. Phys. A: Math. Gen. \textbf{21} (1988) 2559.

\noindent [7] A. Kratzer, Z. Phys. \textbf{3}:5 (1920) 289.

\noindent [8] P. M. Morse, Phys. Rev. \textbf{34}:1 (1929) 57.

\noindent [9] C. Eckart, Phys. Rev. \textbf{35} (1930) 1303.

\noindent [10] M. F. Manning, N. Rosen, Phys. Rev. \textbf{44} (1933) 951.

\noindent [11] G. P\"{o}schl and E. Teller Z. Phys. \textbf{83}:(3-4) (1933) 143.

\noindent [12] L. Hulth\`{e}n, Ark. Mat. Astron. Fys. B \textbf{29}  (1942)  1.

\noindent [13] D. S. Saxon and R.D. Woods, Phys.Rev. \textbf{95} (1954) 577.

\noindent [14] A. A. Makarov, Ya. A. Smorodinsky, Kh.Valiev, P. Winternitz, and Nuovo

\noindent       Cimento A.\textbf{52} (1967) 1061.

\noindent [15] H. Hartmann, Theor. Chim. Acta \textbf{24} (1972) 201.

\noindent [16] A. Hautot, J. Math.Phys. \textbf{14} (1973) 1320.

\noindent [17] F. Cooper, A. Khare, and U. Sukhatme, Supersymmetry in Quantum Mechnics,

\noindent      World Scientific, 2001.

\noindent [18] F. Cooper, A. Khare, U. Sukhatme, Phys. Rep.  \textbf{251} (1995) 267.

\noindent [19] S.-H. Dong, Factorization Method in Quantum Mechanics, Springer, Dordrecht, 2007.

\noindent [20] J. L. Schiff, The Laplace Transform: Theory and Applications, Springer, New York, 1999.

\noindent [21] J. Cai, P. Cai, and A. Inomata, Phys. Rev. A \textbf{34} (1986) 4621.

\noindent [22] A. Z. Tang and F. T. Chan, Phys. Rev. A \textbf{35} (1987) 911.

\noindent [23] B. Roy and R. Roychoudhury, J. Phys. A: Math. Gen. \textbf{20} (1987) 3051.

\noindent [24] F. Dominguez-Adame, Phys. Lett. A \textbf{136} (1989) 175.

\noindent [25] G. Chen, Z. D. Chen, and Z. M. Lou, Phys. Lett. A \textbf{331} (2004) 374.

\noindent [26] B. Talukar, A. Yunus, and M. R. Amin, Phys. Lett. A \textbf{141} (1989) 326.

\noindent [27] L. Chetouani, L. Guechi, A. Lecheheb, T. F. Hammann, and A. Messouber,

\noindent       Physica A \textbf{234} (1996) 529.

\noindent [28] I. B. Okon, O. Popoola, and C. N. Isonguyo. Advances in High Energy Physics Vol.2017,

\noindent      Article ID 9671816, 24.

\noindent [29] Y.-F. Cheng and T.-Q. Dai, Commun. Theor. Phys., \textbf{48} (2007) 431.

\noindent [30] Chang-Yuan Chen, Dong-Sheng Sun, and Fa-Lin Lu, Phys. Lett. A \textbf{370} (2007) 219.

\noindent [31] Wen-Chao Qiang, Run-Suo Zhou, and Yang Gao, Phys. Lett. A \textbf{371} (2007) 201.

\noindent [32] S.-H. Dong and M. Lozada-Cassou, Phys. Scr. \textbf{74} (2006) 285.

\noindent [33] G.-F. Wei, Z.-Z. Zhen, and S.-H. Dong, Cent. Eur. J. Phys. \textbf{7} (1), (2009) 175.

\noindent [34] Sh. M. Nagiyev and A. I. Ahmadov. Int. J. Mod. Phys. A\textbf{34} (2019) 1950089.

\noindent [35] A. O. Barut and R. Raczka, Theory of group representations and applications,

\noindent       Polish Scientific Publishers, Warszawa, 1977.

\noindent [36] C. C. Gerry, Phys. Let. A \textbf{118}:9 (1986) 445.

\noindent [37] V. Fack, H. De Meyer, and G. V. Berghe, J. Math. Phys. \textbf{27}:5 (1986) 1340.

\noindent [38] P. Matthys and H. De Meyer, Phys. Rev. A \textbf{38}:3 (1988) 1168.

\noindent [39] H. Bateman and A. Erdeìlyi. Higher Transcendental Functions, Vol.2,

\noindent       McGrow-Hill Book Company, New York, 1953.

\noindent [40] B. L. Voronov, D. M. Gitman, A. D. Levin, et al. Theor Math Phys. \textbf{187} (2016) 633.

\noindent [41] Sh. M. Nagiyev, Theor. Math. Phys. \textbf{80} (1989) 40.

\noindent [42] R. Koekoek, P. A. Lesky, and R. F. Swarttow, Hypergeometric orthogonal polynomials
and their q-analogies, Springer-Verlag, Berlin, 2010.

\noindent [43] W.-C. Qiang, Chiness Phys. \textbf{12} (10) (2003) 1054.

\end{document}